\PassOptionsToClass{10pt}{revtex4-2}
\documentclass[aps,prl,showpacs,floatfix,twocolumn,byrevtex,superscriptaddress]{revtex4-1}

\usepackage{amsmath}
\usepackage{amssymb}
\usepackage{amstext}
\usepackage{amsopn}
\usepackage{amsfonts}
\usepackage{amsxtra}
\usepackage[english]{babel}
\usepackage{graphicx}
\usepackage{bm}
\usepackage{multirow}
\usepackage{dcolumn}
\usepackage{color}
\usepackage{hyperref}
\usepackage{todonotes}
\usepackage{verbatim}
\usepackage{soul}
\usepackage{xcolor}
\begin{document}

\title{The Hund’s Superconductor Li(Fe,Co)As} 

\author{H. Miao}\email[]{miaoh@ornl.gov}
\affiliation{Materials Science and Technology Division, Oak Ridge National Laboratory, Oak Ridge, Tennessee 37831, USA}
\author{Y. L. Wang}
\affiliation{Condensed Matter Physics and Materials Science Department, Brookhaven National Laboratory, Upton, New York 11973, USA}
\author{J.-X. Yin}
\affiliation{Laboratory for Topological Quantum Matter and Advanced Spectroscopy (B7), Department of Physics, Princeton, New Jersey 08544, USA}
\author{J. Zhang}
\affiliation{Materials Science and Technology Division, Oak Ridge National Laboratory, Oak Ridge, Tennessee 37831, USA}
\author{S. Zhang}
\affiliation{Laboratory for Topological Quantum Matter and Advanced Spectroscopy (B7), Department of Physics, Princeton, New Jersey 08544, USA}
\author{M. Z. Hasan}
\affiliation{Laboratory for Topological Quantum Matter and Advanced Spectroscopy (B7), Department of Physics, Princeton, New Jersey 08544, USA}
\author{R. Yang}
\affiliation{Laboratorium f$\ddot{u}$r Festk$\ddot{o}$rphysik ETH-Z$\ddot{u}$rich, 8093 Z$\ddot{u}$rich, Switzerland}
\author{X. C. Wang}
\affiliation{Beijing National Laboratory for Condensed Matter Physics, Institute of Physics, Chinese Academy of Sciences, Beijing 100190, China}
\affiliation{School of Physical Sciences, University of Chinese Academy of Sciences, Beijing 100190, China}
\author{C. Q. Jin}
\affiliation{Beijing National Laboratory for Condensed Matter Physics, Institute of Physics, Chinese Academy of Sciences, Beijing 100190, China}
\affiliation{School of Physical Sciences, University of Chinese Academy of Sciences, Beijing 100190, China}
\author{T. Qian}
\affiliation{Beijing National Laboratory for Condensed Matter Physics, Institute of Physics, Chinese Academy of Sciences, Beijing 100190, China}
\affiliation{School of Physical Sciences, University of Chinese Academy of Sciences, Beijing 100190, China}
\author{H. Ding}
\affiliation{Beijing National Laboratory for Condensed Matter Physics, Institute of Physics, Chinese Academy of Sciences, Beijing 100190, China}
\affiliation{School of Physical Sciences, University of Chinese Academy of Sciences, Beijing 100190, China}
\author{H.-L. Lee}
\affiliation{Materials Science and Technology Division, Oak Ridge National Laboratory, Oak Ridge, Tennessee 37831, USA}
\author{G. Kotliar}
\affiliation{Physics and Astronomy Department, Rutgers University, Piscataway, New Jersey 08854, USA}
\affiliation{Condensed Matter Physics and Materials Science Department, Brookhaven National Laboratory, Upton, New York 11973, USA}

\date{\today}


\begin{abstract}


We combine transport, angle-resolved photoemission spectroscopy and scanning tunneling spectroscopy to investigate several low energy manifestations of the Hund coupling  in a canonical FeSC family Li(Fe,Co)As. We determine the doping dependence of the   coherent-incoherent crossover temperature and  the quasi-particle effective mass enhancement in the normal state.  Our tunnelling spectroscopy result in the superconducting state supports the idea that superconductivity emerging from Hund’s metal state displays a universal maximal superconducting gap vs transition temperature (2$\Delta_{max}/k_{B}T_{c}$) value, which is independent of doping level and $T_{c}$.

\end{abstract}

\maketitle

A decade after the first discovery of the high-$T_c$ iron-based superconductors (FeSCs), the pairing mechanism and the interplay between the normal and superconducting state in these materials remain elusive \cite{Kamihara2008,Hirschfeld2011, Scalapino2012}. While the phase diagram of the FeSCs shares some similarities with the cuprate high-$T_c$ superconductors, FeSCs are multi-orbital semimetals rather than doped Mott insulators with an effective single orbital \cite{Zhang1988}. The increased orbital degree of freedom is found to qualitatively change the underlying electronic dynamics \cite{Takahiro2012,Hiroaki2010}. In particular, dynamical mean-field theory (DMFT) studies have shown that strong electron-electron correlations in the FeSCs originate from the inter-orbital Hund’s coupling, $J_H$, rather than the on-site intra-orbital Coulomb interaction, $U$ \cite{Werner2008,Haule2009,Yin2011,Yin2011b,Werner2012,Georges2013, Medici2014,Stadler2015,Walter2020}. Consequently, FeSCs are often dubbed as Hund’s metals \cite{Yin2011,Georges2013}.

Evidences of Hund’s metals have been found in the paramagnetic state (PM) of several FeSC families, including doping dependent effective mass \cite{Medici2014}, and orbital differentiations \cite{Yi2013,Miao2016,Yin2014,Li2016}. More recently, theoretical analysis of the Hundness on superconductivity is linked to the ratio 2$\Delta_{max}/k_{B}T_{c}$ in FeSCs \cite{miao2018,Lee2018}, where  $\Delta_{max}$ is the largest superconducting gap in the momentum space. Despite these important experimental and theoretical progresses, justification of Hund’s superconductivity requires a synergetic transport and spectroscopy studies of the Hundness that covers the whole superconducting dome as well as their normal state. Here, using transport, angle-resolved photoemission spectroscopy (ARPES), scanning tunneling spectroscopy (STM) and numerical calculations, we demonstrate that the canonical FeSC, Li(Fe,Co)As, is a Hund’s superconductor, where the Hundness controls the entire temperature ($T$) vs doping ($x$) phase diagram. Our results place strong experimental evidence to support a pairing mechanism driven by the Hundness-induced critical spin fluctuations for the FeSCs \cite{Lee2018}.

%
\begin{figure*}[tb]
\includegraphics[width=16 cm]{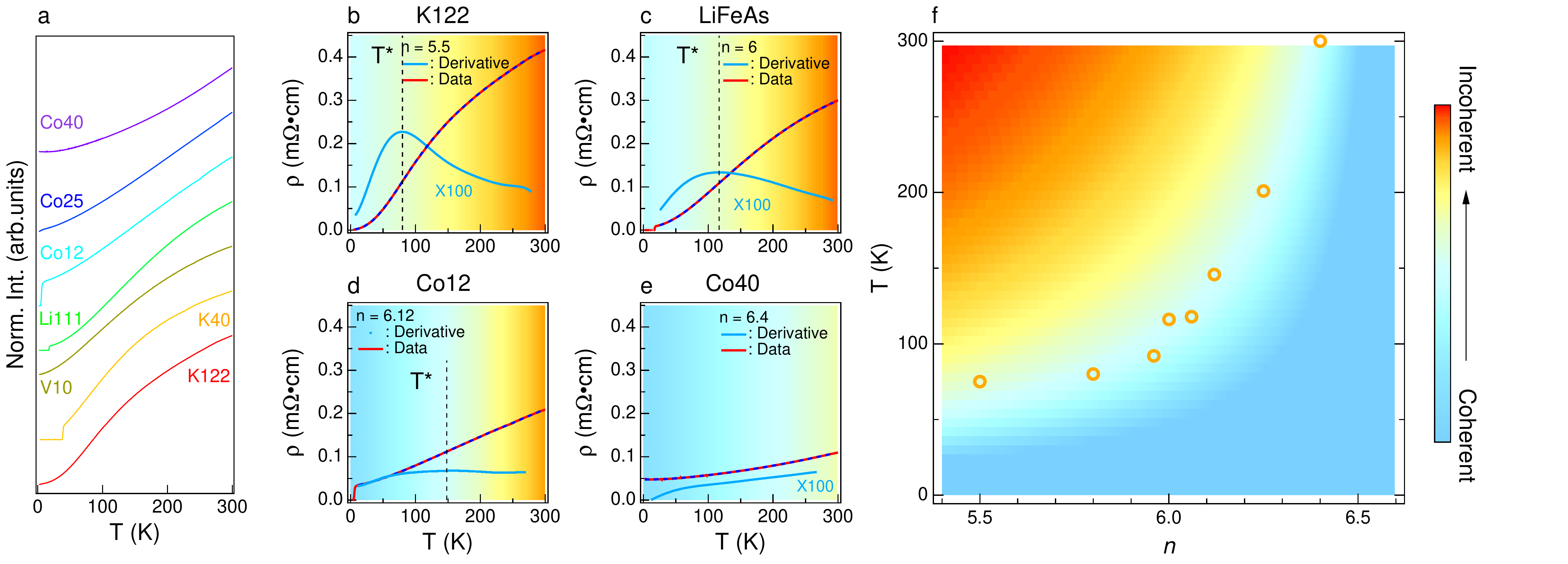}
\caption{Doping dependent coherent-incoherent crossover. (a) Normalized resistivity of LiFe$_{0.6}$Co$_{0.4}$As (Co40), LiFe$_{0.75}$Co$_{0.25}$As (Co25), LiFe$_{0.88}$Co$_{0.12}$As (Co12), LiFeAs (Li111), LiFe$_{0.9}$V$_{0.1}$As (V10), Ba$_{0.6}$K$_{0.4}$Fe$_2$As$_2$ (K40) and KFe$_2$As$_2$ (K122). Resistivity (red curve) of K122, LiFeAs, Co12 and Co40 are shown in (b)-(e). Dashed lines are polynomial fittings, $\rho_{FIT}(T)=\sum_{n=0}^{9}a_{n}T^{n}$, of the resistivity. Cyan curves in (b)-(e) are the first derivative ($\frac{d\rho_{FIT}}{dT}$) of the fitted resistivity. The dashed vertical line marks the maximum of $\frac{d\rho_{FIT}}{dT}$, which is defined as the coherent-incoherent crossover temperature, $T^*$ \cite{Haule2009}. (f) Extracted $n$-dependent $T^*$ is remarkably consistent with theoretical predictions for Hund’s metal \cite{Haule2009,Werner2008,Werner2012}. Here $n$ is converted from the Co concentration x (\%) \cite{Supp}. The background colors in (b)-(f) represent the electronic coherence determined by the resisitivity curvature.}
\label{Fig1}
\end{figure*}

We choose Li(Fe,Co)As for this study as its phase diagram is not intervened by any magnetic or nematic long range orders \cite{Supp}. This condition is crucial as the Hund’s metal physics is essentially a description of the PM phase. Another reason to study Li(Fe,Co)As is that the Hund’s coupling induced characteristic orbital differentiation in the dynamical charge and magnetic excitations are observed in the parent compound LiFeAs \cite{Miao2016,Yin2014, Li2016}. 

We first explore the PM state of electron and hole doped LiFeAs via transport measurements (see ref.~\cite{Supp} for experimental and computational methods). Figure~\ref{Fig1}a shows the normalized resistivity of Li(Fe,TM)As (TM=V, Co) that cover a wide doping range, corresponding to the occupation number, $n$, from 5.97 to 6.4 \cite{Shi2017,Supp}. For comparison, we also include the normalized resistivity of Ba$_{0.6}$K$_{0.4}$Fe$_2$As$_2$ (K40) and KFe$_2$As$_2$ (K122) that correspond to $n$ = 5.8 and 5.5, respectively \cite{Ding2008,Fang2015}. From the electron-doped side to the hole-doped side, the resistivity curvature undergoes a qualitative change: it remains positive in the entire temperature range in LiFe$_{0.6}$Co$_{0.4}$As (Co40) but changes sign at high temperature in LiFeAs. Previous DMFT studies have shown that the curvature change in resistivity represents a coherent-incoherent crossover of the electronic system \cite{Haule2009}. To quantify this change, we show the resistivity of K122, LiFeAs, LiFe$_{0.88}$Co$_{0.12}$As (Co12) and Co40 in Fig.~\ref{Fig1}b-e. First of all, we notice that the room temperature resistivity $\rho_{300}$ decreases by a factor of four from K122 to Co40 despite a larger number of impurities in Co40, suggesting reduced electronic interaction via electron doping. We then fit the resistivity curves above $T_c$ with 9th order polynomial and use the maximum of the first derivative to characterize the coherent-incoherent crossover temperature, $T^*$ (Fig.~\ref{Fig1}b-e). Figure~\ref{Fig1}f summarizes $T^*$ as function of $n$, where the $T^*$ increases from 80 K in K122 to room temperature in Co40, remarkably consistent with previous LDA+DMFT calculations \cite{Haule2009,Werner2012} and in agreement with our resistivity calculations for LiFeAs (see supplementary materials \cite{Supp}). 

%
\begin{figure*}[tb]
\includegraphics[width=13 cm]{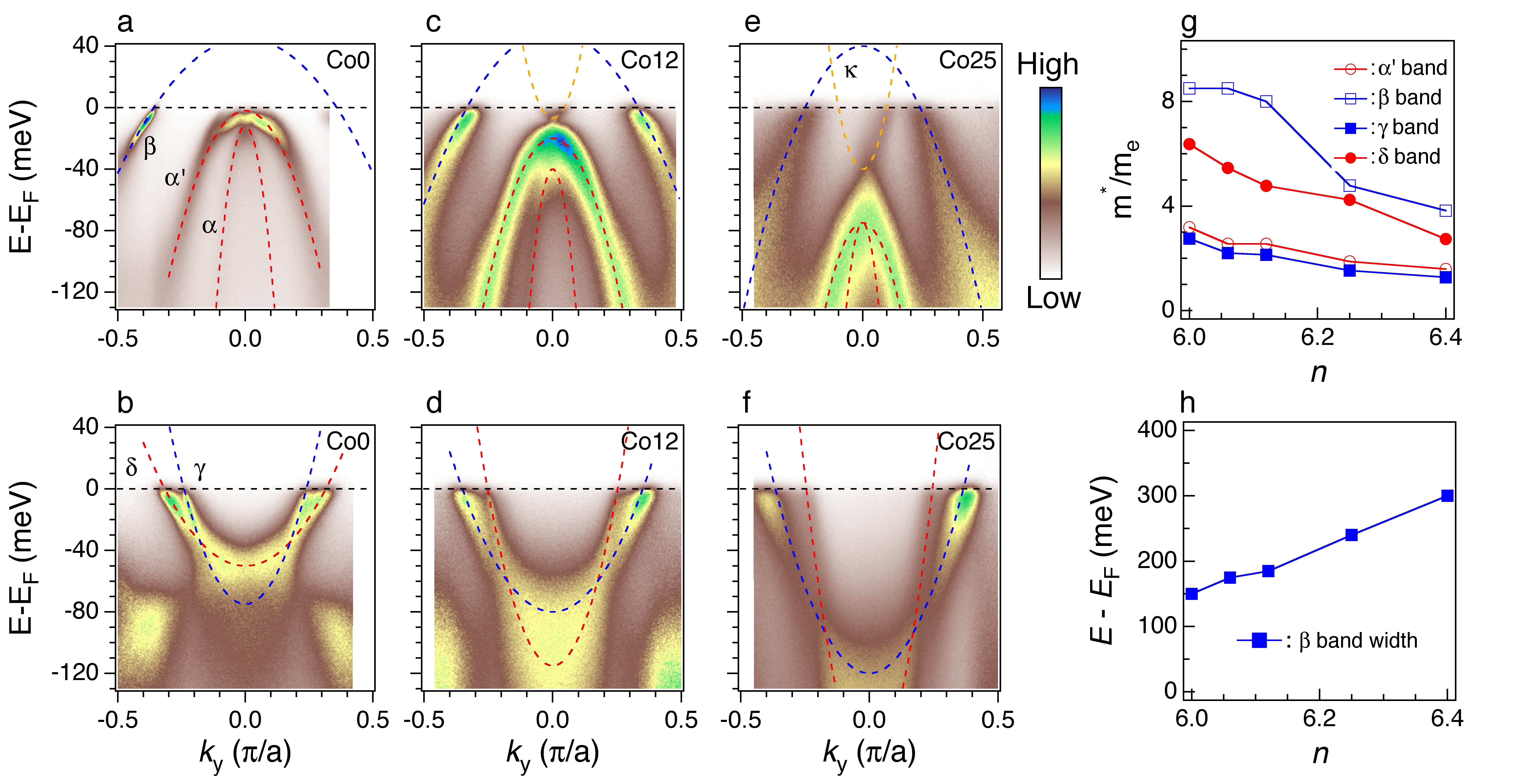}
\caption{Doping dependent band structure of Li(Fe,Co)As. The ARPES intensity plot near the $\Gamma$ (a-c) and M (d-f) point. Dashed lines are parabolic fitting of the band dispersion (peak intensity). The extracted doping dependent effective mass of the $\alpha'$, $\beta$, $\gamma$ and $\delta$ band are shown in (g). (h) shows the $n$-dependent bandwidth of $\beta$ band, determined by the band top of the $\beta$ band and band bottom of the $\gamma$ band. Since the $\beta$ and $\gamma$ band are degenerate at the M point, the energy difference between the top of $\beta$ and the bottom of $\gamma$g is equivalent to the band width of the $\beta$ band, which is broadened by a factor of two from LiFeAs to Co40, consistent with the doping dependent effective mass of $\beta$ band in (g).}
\label{Fig2}
\end{figure*}

The evolution of coherent-incoherent crossover in resistivity mirrors the $n$-dependent electronic interactions. To directly show this trend, we plot the doping-dependent electronic structure measured by ARPES at the $\Gamma$ and M point in Fig.~\ref{Fig2}a-f. The dashed lines are parabolic fittings of the band structure with their colors representing the main orbital characters. Figure~\ref{Fig2}g summarizes the extracted band and orbital-resolved effective mass, $m^*$, as function of $n$. We find that $m^*$ of all bands are reduced by a factor of two from LiFeAs to Co40. This is in agreement with the bandwidth evolution of the $\beta$ band (Fig.~\ref{Fig2}h), which is determined by the energy difference between the top of the $\beta$ band and the bottom of the $\gamma$ band. 

\begin{figure*}[tb]
\includegraphics[width=14 cm]{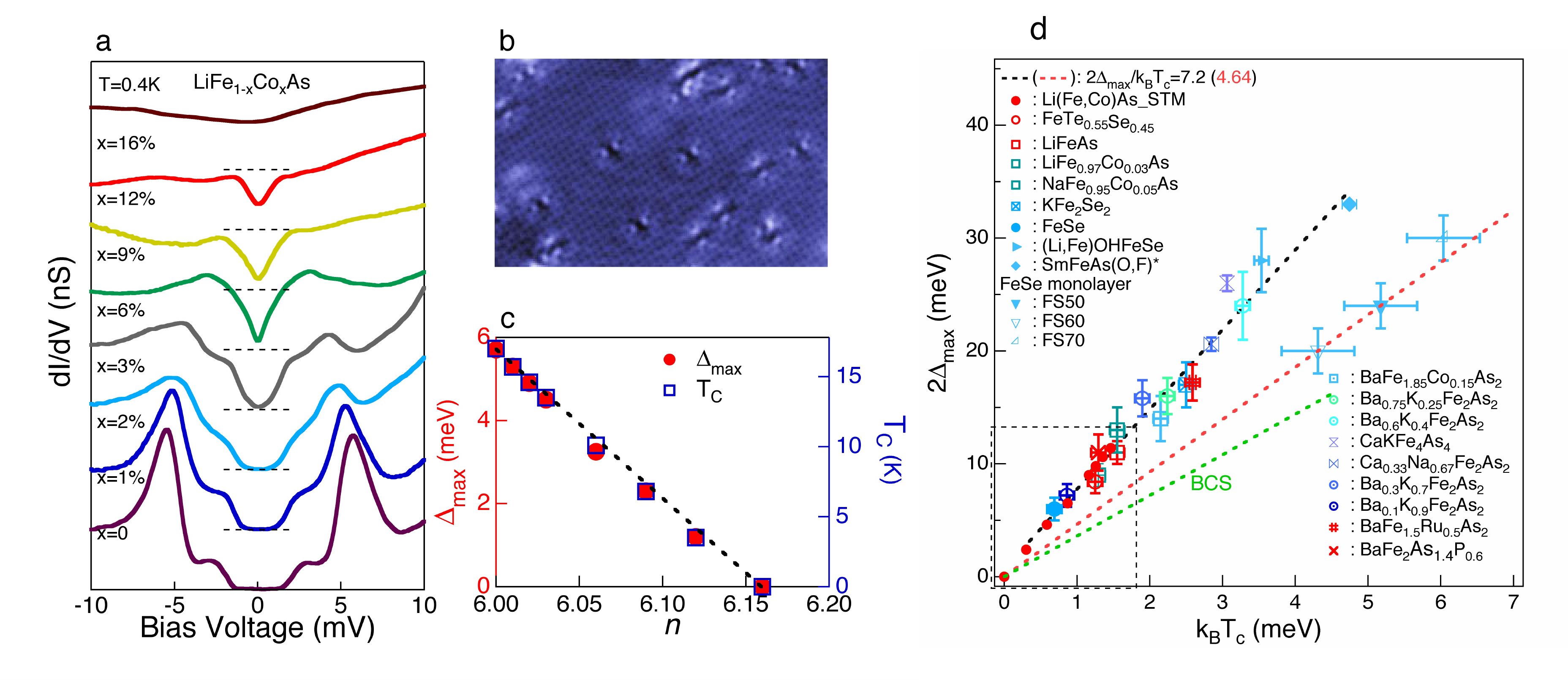}
\caption{Doping-dependent SC gap and universal 2$\Delta_{max}/k_{B}T_{c}$ scaling. (a) Averaged STS spectra measured at $T$=400~mK. The dashed lines centered the zero bias voltage represent zero $dI/dV$. (b) The STM topography of LiFe$_{0.99}$Co$_{0.01}$As. (c) The $T_c$ and extracted $\Delta_{max}$ as function of doping are shown in red circles and blue squares. Data shown in (a) and (c) are adapted from ref.~\cite{Yin2019}. (d) shows the universal 2$\Delta_{max}/k_{B}T_{c}$ scaling of FeSCs \cite{miao2018}. The black and green dashed lines correspond to Hund’s superconductor and BCS superconductor, respectively. The red dashed line is a fit to 2$\Delta_{max}/k_{B}T_{c}$ extracted from FeSe monolayer.}
\label{Fig3}
\end{figure*}

The profound doping-dependent $T^*$ and $m^*$ strongly indicate that the entire superconducting dome ($6<n<6.16$) of Li(Fe,Co)As is emerging from the Hund’s metal normal state. To uncover the Hundness in the superconducting state, we extract the doping-dependent SC gap from scanning tunneling microscopy/spectroscopy (STM/STS) measurements of Li(Fe,Co)As at 400 mK, deep in the superconducting state \cite{Yin2019}. Figure~\ref{Fig3}b shows a typical atomically resolved STM topography image of LiFe$_{0.99}$Co$_{0.01}$As. The spatially averaged STS spectra of eight different doping levels with $T_c$ ranging from 18 to 4~K are shown in Fig.\ref{Fig3}a. The dashed lines mark the zero-intensity value for each doping. The extracted $\Delta_{max}$ as a function of doping is plotted on top of the $T_c$ vs $n$ diagram in Fig.~\ref{Fig3}c. The perfect overlap of these two plots demonstrates a universal 2$\Delta_{max}/k_{B}T_{c}\sim7.7$ scaling in the entire Li(Fe,Co)As superconducting phase. This value is nicely consistent with recent model study, where a universal 2$\Delta_{max}/k_{B}T_{c}\sim7.2$ is derived by assuming the Cooper pairs are “glued” by the Hundness induced local spin fluctuations with a characteristic $\chi_{sp}^{''}(\omega)\propto\omega^{-6/5}$ scaling \cite{Lee2018,Supp,Wang2020}. $\chi_{sp}^{''}(\omega)$ is the imaginary part of the local dynamical susceptibility. The nearly perfect agreement between the theory and experiment in the entire phase diagram firmly establishes the Hund’s superconductivity in Li(Fe,Co)As.

%

Our results have significant implications on the pairing mechanism of FeSCs. Figure~\ref{Fig3}d shows the 2$\Delta_{max}$ vs $k_{B}T_{c}$ plot of a large number of FeSCs including Li(Fe,Co)As. The ubiquitous 2$\Delta_{max}/k_{B}T_{c}$ scaling over a large set of FeSCs suggests a common pairing mechanism for all FeSCs. The consistent 2$\Delta_{max}/k_{B}T_{c}$ value between the experimental data and model calculation strongly indicates that the leading pairing interaction is the Hundness-induced local spin fluctuations. This conclusion explains early experimental observations where the SC gap function $\Delta(k)$ follows simple local form factors and involves electronic state away from the Fermi energy $E_F$ \cite{Miao2012,Hu2012,Miao2015}. It should be noted, however, the theoretically derived 2$\Delta_{max}/k_{B}T_{c}$ is based on a simplified model that neglects several material specific details such as the proximity to electronic nematicity in Ba(Fe,Co)$_2$As$_2$ and bulk FeSe \cite{Yi2011, Fernandes2014, Sprau2017} and strong electron-phonon coupling in FeSe monolayer \cite{Lee2014}. In addition, the presence of non-local interaction can induce moderate gap variations on the Fermi surface \cite{Umezawa2012,Borisenko2010,Bhattacharyya2020}. However, the large number of FeSCs with a similar 2$\Delta_{max}/k_{B}T_{c}$ ratio suggests that the Hundness is robust even in the presence of various low-energy modifications. 

In summary, using transport and spectroscopic techniques, we synergistically uncovered signatures of Hundness from the normal to SC state of Li(Fe,Co)As. Our result establishes the  Hund’s superconductivity in canonical FeSC Li(Fe,Co)As and indicates that the Hundness-induced critical spin fluctuation is the leading pairing interaction in FeSCs. 

This research at Oak Ridge National Laboratory (ORNL) was sponsored by the Laboratory Directed Research and Development Program of ORNL, managed by UT-Battelle, LLC, for the U. S. Department of Energy (ARPES) and by the U.S. Department of Energy, Office of Science, Basic Energy Sciences, Materials Sciences and Engineering Division (transport). Y.W. was  supported by the US Department of energy, Office of Science, Basic Energy Sciences as a part of the Computational Materials Science Program through the Center for Computational Design of Functional Strongly Correlated Materials and Theoretical Spectroscopy.  GK was supported by NSF DMR-1733071 Work at the IOP is supported by grants from the National Natural Science Foundation of China (11888101, 11674371), and the Ministry of Science and Technology of China (2016YFA0401000). Work at Princeton University was supported by the Gordon and Betty Moore Foundation (GBMF4547/ Hasan) and the United States Department of energy (US DOE) under the Basic Energy Sciences program (Grant No. DOE/BES DE- FG-02-05ER46200). H.M., Y.L.W and J.-X.Y are contributed equally to this work.


\bibliography{ref}

\end{document}